\documentclass[aps,prl,showpacs,twocolumn,amsmath,amssymb]{revtex4}
\usepackage[english]{babel}
\usepackage{latexsym}
\usepackage{graphics}
\usepackage{epsfig}

\begin{document}

\title{Orientation of Nd$^{3+}$ dipoles in yttrium aluminium garnet~:\\
a simple yet accurate model}

\author{Sylvain~Schwartz$^{1}$,}
\email{sylvain.schwartz@thalesgroup.com} \author {Gilles~Feugnet$^1$, Maxence Rebut$^1$, Fabien Bretenaker$^2$, and
Jean-Paul~Pocholle$^1$}

\affiliation{$^1$Thales Research and Technology France, Campus Polytechnique, 1 avenue Augustin Fresnel, F-91767 Palaiseau Cedex, France\\
$^2$Laboratoire Aim\'e Cotton, CNRS - Universit\'e Paris Sud 11, Campus d'Orsay, F-91405 Orsay Cedex, France}

\date{\today}


\begin{abstract}
We report an experimental study of the 1.064~$\mu$m transition dipoles in neodymium doped yttrium aluminium garnet
(Nd-YAG) by measuring the coupling constant between two orthogonal modes of a laser cavity for different cuts of the YAG
gain crystal. We propose a theoretical model in which the transition dipoles, slightly elliptic, are oriented along the
crystallographic axes. Our experimental measurements show a very good quantitative agreement with this model, and
predict a dipole ellipticity between 2\% and 3\%. This work provides an experimental evidence for the simple description
in which transition dipoles and crystallographic axes are collinear in Nd-YAG (with an accuracy better than 1~deg), a
point that has been discussed for years.
\end{abstract}

\maketitle


While Nd-YAG is one of the most (if not the most) commonly used solid-state laser crystals, the exact orientation of
transition dipoles within it is still, paradoxically, an unresolved problem. This is probably owing to the fact that for
most applications it is sufficient to consider the Nd-YAG crystal (usually grown along the $\langle 111 \rangle$
crystallographic axis) as isotropic, although it has been known for long \cite{dillon, ziel} that Nd$^{3+}$ ions in this
configuration rather see a $D_2$ symmetry, with six possible dodecahedral orientations. The influence of crystal
symmetry on dipole orientations has been previously studied in saturable absorbers such as Cr-YAG
\cite{apl_eilers,brignon,pra_brunel} and Tm-YAG \cite{pra_tm,prb_tm}. In the first case, it has been clearly established
that transition dipoles were aligned with the crystal axes (labeled $\langle 100 \rangle$, $\langle 010 \rangle$ and
$\langle 001 \rangle$) \cite{brignon,pra_brunel} while in the second case it has been found that they were rather
collinear with the $\langle 110 \rangle$, $\langle 011 \rangle$ and $\langle 101 \rangle$ directions \cite{prb_tm}. In
the case of Nd-YAG, the answer to this question is still unclear in spite of several previous studies involving in
particular dynamical polarization effects in Nd-YAG lasers \cite{spie,slovenes,lar,mckay}.

In this paper, we propose a new approach to probe the orientation of transition dipoles in Nd-YAG, by measuring the
coupling constant between two linearly-polarized orthogonal modes of a laser cavity for different cuts of the gain
crystal, using a steady-state method similar to the one described in \cite{apl_brunel}. The measured coupling constant
is a dimensionless ratio between cross-saturation and self-saturation coefficients, which is relatively independent of
most laser parameters (pumping rate, birefringence,$\ldots$), hence a good indicator for testing the validity of
theoretical models. Our study deals for the most part with the 1064.15~nm emission line, sometimes referred to as R2,
between the upper doublet of $^4 F _{3/2}$ and the Y3 level of $^4 I_{11/2}$. As a matter of fact, it is known from
previous studies \cite{kushida,weber} that the R1 line at 1064.4~nm (between the lower doublet of $^4 F_{3/2}$ and the
Y2 level of $^4 I_{11/2}$) has a very small contribution to the overall gain, especially at low pumping rates.

The paper is organized as follows. We first propose a theoretical model for calculating the coupling constant between
two orthogonal modes of a Nd-YAG laser cavity, on the assumption that transition dipoles are collinear with YAG
crystallographic axes. Starting from the very simple case of linear transition dipoles, we then generalize the model to
the case of (slightly) elliptical dipoles, showing a very good agreement with the experimental value of the coupling
constant published in \cite{apl_brunel} with $\langle 111 \rangle$-cut Nd-YAG. We then apply this model to the
description of our own experimental configuration, namely a two-mode laser cavity using as a gain medium a Nd-YAG
crystal either $\langle 111 \rangle$ or $\langle 100 \rangle$-cut. In both cases, the measured value of the coupling
constant is compared with the corresponding theoretical prediction. The results, implications and perspectives of this
work are finally discussed.

\section{Expression of the coupling constant in the case of linear dipoles}

In the usual semiclassical description of lasers, the transition dipole is often modeled by an operator
$\hat{\mathbf{d}}$ associated with the linear vector $\mathbf{d}=d\mathbf{u}$, whose interaction with a
linearly-polarized electric field $\mathbf{E}=E\mathbf{x}$ is described quantum mechanically by the dipolar Hamiltonian
$-\hat{\mathbf{d}}\cdot\mathbf{E}$. Assuming furthermore that dipole coherence lifetime is much shorter than photon
lifetime $\tau$ and population inversion lifetime $T_1$ (which is indeed the case in Nd-YAG lasers \cite{Siegman}), the
dipolar interaction can be described by the following terms in the rate equations for the population inversion density
$N$ and field intensity $I$ \cite{Siegman}~:
\begin{equation} \label{sys1}
\left\{ \begin{split} & \left. \frac{\textrm{d}N}{\textrm{d}t}\right|_\textrm{int}=-\frac{N}{T_1} \, \frac{I}{I_s}
\cos^2 \left(\widehat{\mathbf{x},\mathbf{u}} \right) \;,\\
& \left. \frac{\textrm{d}I}{\textrm{d}t}\right|_\textrm{int}= \sigma c N I \cos^2 \left(\widehat{\mathbf{x},\mathbf{u}}
\right) \;, \end{split} \right. \,
\end{equation}
where $I_s$ is the saturation intensity, $\sigma$ the interaction cross section, $c$ the speed of light in vacuum, and
where $\mathbf{u}$ and $\mathbf{x}$ are assumed to be unit vectors. The overall rate equations for $I$ and $N$ in a
simple Lamb's laser model then read \cite{Siegman}~:
\begin{equation} \label{sys2}
\frac{\textrm{d}N}{\textrm{d}t}=W-\frac{N}{T_1} + \left. \frac{\textrm{d}N}{\textrm{d}t}\right|_\textrm{int} \quad
\textrm{and} \quad \frac{\textrm{d}I}{\textrm{d}t}= -\frac{I}{\tau}+ \left.
\frac{\textrm{d}I}{\textrm{d}t}\right|_\textrm{int},
\end{equation}
where $W$ is the pumping rate. As regards Nd-YAG lasers, the latter equations are sufficient to describe satisfactorily
most of the experimentally observed phenomena such as relaxation oscillations \cite{relax} or spiking during laser
turn-on \cite{spiking}. However, they do not allow an accurate description of mode coupling in a Nd-YAG laser cavity
\cite{apl_brunel}.

\subsection{General expression of the coupling constant}

To this end, one must take into account, in addition with the two laser modes, the existence of several possible
orientations for the transition dipoles. In the following model, we shall assume three possible orientations
corresponding to the unitary vectors $\mathbf{u}_1$, $\mathbf{u}_2$ and $\mathbf{u}_3$, and associated with population
inversion densities $N_1$, $N_2$ and $N_3$. We furthermore consider, in keeping with \cite{apl_brunel} and with the
experiment described later on in this paper, the case of a laser with two modes linearly polarized along the
$\mathbf{x}_1$ and $\mathbf{x}_2$ unitary vectors, and associated with intensities $I_1$ and $I_2$. The semiclassical
equations for the dipolar interaction (\ref{sys1}) can be generalized as follows~:
\begin{equation} \label{sysys1}
\left\{ \begin{split} & \left. \frac{\textrm{d}N_i}{\textrm{d}t}\right|_\textrm{int}=-\frac{N_i}{T_1} \sum_{j=1,2}
\frac{I_j}{I_s^j} \, \cos^2
\left(\widehat{\mathbf{x}_j,\mathbf{u}_i} \right) \;,\\
& \left. \frac{\textrm{d}I_j}{\textrm{d}t}\right|_\textrm{int}= \sigma c I_j \sum_{i=1}^3 N_i \cos^2
\left(\widehat{\mathbf{x}_j,\mathbf{u}_i} \right) \;, \end{split} \right. \,
\end{equation}
where we have introduced two possibly different saturation intensities $I_s^1$ and $I_s^2$. The overall rate equations
(\ref{sys2}) for $N_i$ and $I_j$ become in this case~:
\begin{equation} \label{syssys2}
\frac{\textrm{d}N_i}{\textrm{d}t}=W-\frac{N_i}{T_1} + \left. \frac{\textrm{d}N_i}{\textrm{d}t}\right|_\textrm{int}
\textrm{and} \;\; \frac{\textrm{d}I_j}{\textrm{d}t}= -\frac{I_j}{\tau_j}+ \left.
\frac{\textrm{d}I_j}{\textrm{d}t}\right|_\textrm{int},
\end{equation}
where we have introduced different loss coefficients for each mode $\gamma_j=1/\tau_j$. In equations (\ref{syssys2}),
$W$ has been chosen to be independent of $i$, which corresponds to the case of isotropic pumping (this has been checked
experimentally, see further in this paper). In the steady-state regime and for near-threshold operation (i.e. $I_j/I_s^j
\ll 1$), equations (\ref{sysys1}) and (\ref{syssys2}) can be rewritten as~:
\begin{equation} \label{steadyst1}
 N_i=W T_1 \left[1- \sum_{j=1,2} \frac{I_j}{I_s^j} \, \cos^2
\left(\widehat{\mathbf{x}_j,\mathbf{u}_i} \right) \right] \;,
\end{equation}
and~:
\begin{equation} \label{steadyst2}
\gamma_j = \sigma c \sum_{i=1}^3 N_i \cos^2 \left(\widehat{\mathbf{x}_j,\mathbf{u}_i} \right) \;.
\end{equation}
The coupling constant $C$, initially defined by Lamb as the ratio between cross-saturation coefficients and
self-saturation coefficients \cite{lamb} in the case of lasers with short population inversion lifetime ($T_1 \ll
\tau$), can be generalized to other kinds of lasers like Nd-YAG by the following (more general) definition, involving
small variations from the steady-state regime \cite{apl_brunel}~:
\begin{equation} \label{def}
C = \frac{(\Delta I_1/\Delta \gamma_2)(\Delta I_2/\Delta \gamma_1)}{(\Delta I_1/\Delta \gamma_1)(\Delta I_2/\Delta
\gamma_2)} \;.
\end{equation}
Using this definition, a straightforward calculation starting from equations (\ref{steadyst2}) and using equations
(\ref{steadyst1}) for the expression of $N_i$ leads to the following formula~:
\begin{equation} \label{C}
C = \frac{\left(\displaystyle{\sum_{i=1}^3} \cos^2 \left(\widehat{\mathbf{x}_1,\mathbf{u}_i} \right) \cos^2
\left(\widehat{\mathbf{x}_2,\mathbf{u}_i} \right) \right)^2}{\displaystyle{\sum_{i=1}^3} \cos^4
\left(\widehat{\mathbf{x}_1,\mathbf{u}_i} \right) \displaystyle{\sum_{i=1}^3} \cos^4
\left(\widehat{\mathbf{x}_2,\mathbf{u}_i} \right)} \;.
\end{equation}
One important point is that this expression depends only on the overall geometry, making the coupling constant a useful
tool for studying dipoles orientation.

\subsection{Application to the case of the $\langle 111 \rangle$ crystal}

The case of the $\langle 111 \rangle$ crystal is by far the most common for Nd-YAG lasers. In this configuration, the
laser wave-vector $\mathbf{k}$ is along the $\langle 111 \rangle$ axis, while the laser electric field lies in the
transverse plane. Let us assume that the two laser modes are linearly polarized along the two following transverse
unitary vectors~:
\begin{equation} \label{basis}
\mathbf{x}_1=\frac{1}{\sqrt{2}} \left( \begin{tabular}{c} 1 \\ -1 \\ 0 \end{tabular} \right) \quad \textrm{and} \quad
\mathbf{x}_2=\frac{1}{\sqrt{6}} \left( \begin{tabular}{c} 1 \\ 1 \\ -2 \end{tabular} \right) \;,
\end{equation}
where the coordinates are expressed in the base of the crystallographic axes. Although this orthogonal base of the
transverse plane has been arbitrarily chosen, it should be pointed out that this choice does not affect the final
expression for the coupling constant (calculation shown in the appendix). Differently speaking, the coupling constant is
independent, in the $\langle 111 \rangle$ case, of the crystal orientation. Following the authors of \cite{slovenes}, we
consider transition dipoles $d\mathbf{u}_1$, $d\mathbf{u}_2$ and $d\mathbf{u}_3$ collinear with the crystal axes,
namely~:
\begin{equation} \label{dipol}
\mathbf{u}_1= \left( \begin{tabular}{c} 1 \\ 0 \\ 0 \end{tabular} \right) \quad\!\!\! \textrm{,} \quad\! \mathbf{u}_2=
\left(
\begin{tabular}{c} 0 \\ 1 \\ 0 \end{tabular} \right) \quad\! \textrm{and} \quad\! \mathbf{u}_3= \left(
\begin{tabular}{c} 0 \\ 0 \\ 1 \end{tabular} \right) \;.
\end{equation}
This hypothesis will be self-consistently confirmed by our experimental results later on in this paper. Expression
(\ref{C}) immediately leads in this case to the coupling constant value $C=1/9 \simeq 0.11$.\\

This value can be compared with the experimental measurement $C \simeq 0.16 \pm 0.03$ from reference \cite{apl_brunel}.
The discrepancy is attributed to the fact that dipoles from each site are
not perfectly linear, but rather slightly elliptic, as will be described in what follows.\\

\section{Taking into account cross-couplings between dipoles from different crystal sites}

\begin{figure}
\begin{center}
\includegraphics[scale=0.55]{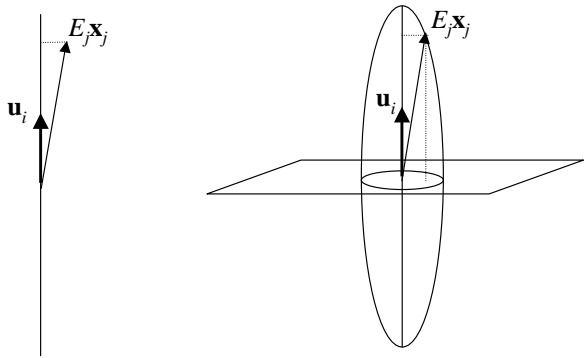}
\end{center}
\caption{Comparison between usual linear dipole coupling (left) and elliptical dipole coupling (right). The latter has
been phenomenologically introduced to allow for cross-coupling between dipoles from different sites.}\label{figure
ellipse}
\end{figure}

In order to account for possible cross-couplings between dipoles from different crystal sites, we shall assume a small
dipole ellipticity $\beta \ll 1$. In this phenomenological description, illustrated on figure \ref{figure ellipse}, the
field-dipole interaction (\ref{sysys1}) takes the new following form~:
\begin{equation} \nonumber
\left\{ \begin{split} & \left. \frac{\textrm{d}N_i}{\textrm{d}t}\right|_\textrm{int}\!=-\frac{N_i}{T_1} \sum_{j=1}^2
\frac{I_j}{I_s^j} \left[ \cos^2 \left(\widehat{\mathbf{x}_j,\mathbf{u}_i} \right) + \beta \sin^2
\left(\widehat{\mathbf{x}_j,\mathbf{u}_i} \right) \right] ,\\
& \left. \frac{\textrm{d}I_j}{\textrm{d}t}\right|_\textrm{int}\!= \sigma c I_j \sum_{i=1}^3 N_i \left[ \cos^2
\left(\widehat{\mathbf{x}_j,\mathbf{u}_i} \right) + \beta \sin^2 \left(\widehat{\mathbf{x}_j,\mathbf{u}_i} \right)
\right] . \end{split} \right.
\end{equation}
This model will be used to calculate a new expression for the coupling constant in both the $\langle 111\rangle$ and
$\langle 100 \rangle$ cases.

\subsection{Case of the $\langle 111 \rangle$ crystal}

The latter equations can be applied to the previously-studied case of the $\langle 111 \rangle$ Nd-YAG crystal, using
the vectors defined in (\ref{basis}) and (\ref{dipol}). This leads to the following rate equations for the population
inversion densities~:
\begin{equation} \nonumber
\left\{ \begin{split} & \left. \frac{\textrm{d}N_1}{\textrm{d}t} \right|_\textrm{int}=
-\frac{N_1}{T_1}\left[\frac{I_1}{I_s^1} \left(\frac{1}{2}+\frac{\beta}{2} \right) +\frac{I_2}{I_s^2} \left(
\frac{1}{6} + \frac{5 \beta}{6} \right) \right] \;, \\
& \left. \frac{\textrm{d}N_2}{\textrm{d}t} \right|_\textrm{int}= -\frac{N_2}{T_1}\left[\frac{I_1}{I_s^1}
\left(\frac{1}{2}+\frac{\beta}{2} \right) +\frac{I_2}{I_s^2} \left(
\frac{1}{6} + \frac{5 \beta}{6} \right) \right] \;, \\
& \left. \frac{\textrm{d}N_3}{\textrm{d}t} \right|_\textrm{int}= -\frac{N_3}{T_1}\left[\frac{I_1}{I_s^1} \beta
+\frac{I_2}{I_s^2} \left( \frac{2}{3} + \frac{\beta}{3} \right) \right] \;,
\end{split} \right.
\end{equation}
and for the laser modes's intensities~:
\begin{equation} \nonumber
\left\{ \begin{split} & \left. \frac{\textrm{d}I_1}{\textrm{d}t} \right|_\textrm{int}= \sigma c \left[\frac{N_1
(1+\beta)}{2}+\frac{N_2 (1+\beta)}{2} + N_3 \beta \right] I_1\;, \\
& \left. \frac{\textrm{d}I_2}{\textrm{d}t} \right|_\textrm{int}= \sigma c \left[\frac{(N_1 + N_2)
(1+5\beta)}{6}+ N_3 \frac{2+\beta}{3} \right] I_2\;. \\
\end{split} \right.
\end{equation}
We eventually obtain the following expression for the coupling constant in the presence of small elliptical dipolar
coupling, up to the first order in $\beta$~:
\begin{equation} \label{Cwithbeta}
C = \frac{1}{9} + \frac{16}{9} \beta \;.
\end{equation}
Based on the experimental result $C \simeq 0.16 \pm 0.03$ of reference \cite{apl_brunel}, the estimate $\beta \simeq
2.75\%$ ($\pm 0.5 \%$) can be deduced from expression (\ref{Cwithbeta}). A possible explanation for this ellipticity id
the existence of arbitrarily-oriented Nd$^{3+}$ ions residing in defect sites of the YAG matrix, that would induce
energy transfer between dipoles from different intrinsic crystal sites.\\

\subsection{Case of the $\langle 100 \rangle$ crystal}

We now turn to the case of a $\langle 100 \rangle$ crystal, on the reasonable assumption that dipole ellipticity $\beta$
is independent of crystal cut. In this new configuration, the laser wave-vector $\mathbf{k}$ is aligned with one
crystallographic axis (say $\mathbf{u}_3$) and the two orthogonal laser modes are linearly polarized along two
transverse axes defined by~:
\begin{equation} \nonumber
\mathbf{x}_1= \left( \begin{tabular}{c} $\cos \alpha$ \\ $\sin \alpha$ \\ 0 \end{tabular} \right) \quad \textrm{and}
\quad \mathbf{x}_2= \left(
\begin{tabular}{c} $-\sin \alpha$ \\ $\cos \alpha$ \\ 0 \end{tabular} \right) \;,
\end{equation}
where the coordinates have been expressed in the base ($\mathbf{u}_1$,$\mathbf{u}_2$,$\mathbf{u}_3$) of the
crystallographic axes. With the latter definition, the polarization directions of the two laser modes $\mathbf{x}_1$ and
$\mathbf{x}_2$ make an angle $\alpha$ with the crystal (or dipoles) axes $\mathbf{u}_1$ and $\mathbf{u}_2$. This leads,
up to the first order in $\beta$, to the following rate equations for the population inversion densities~:
\begin{equation} \nonumber
\left\{ \begin{split} & \left. \frac{\textrm{d}N_1}{\textrm{d}t} \right|_\textrm{int}=
-\frac{N_1}{T_1}\left[\frac{I_1}{I_s^1} A +\frac{I_2}{I_s^2} B \right] \;, \\
& \left. \frac{\textrm{d}N_2}{\textrm{d}t} \right|_\textrm{int}= -\frac{N_2}{T_1}\left[\frac{I_1}{I_s^1} B
+\frac{I_2}{I_s^2} A \right] \;, \\
\end{split} \right.
\end{equation}
where the following notations have been introduced~:
\begin{equation} \label{expdeC}
A= \cos^2 \alpha + \beta \sin^2 \alpha \quad \textrm{and} \quad B=\sin^2 \alpha + \beta \cos^2 \alpha \;.
\end{equation}
Similarly, one gets the following rate equations for the modes's intensities~:
\begin{equation} \nonumber
\left\{ \begin{split} & \left. \frac{\textrm{d}I_1}{\textrm{d}t} \right|_\textrm{int}= \sigma c \left[N_1
A + N_2 B \right] I_1\;, \\
& \left. \frac{\textrm{d}I_2}{\textrm{d}t} \right|_\textrm{int}= \sigma c \left[N_1 B + N_2 A \right] I_2\;. \\
\end{split} \right.
\end{equation}
It should be mentioned that the influence of $N_3$ has been neglected in this analysis since it would involve only terms
on the order of $\beta^2$ or smaller. One finally obtains $C = 4A^2B^2/(A^2+B^2)^2$, which reduces, up to the first
order in $\alpha^2$ and $\beta$, to the following expression~:
\begin{equation} \label{expdeCfinal}
C=4(\alpha^2 + \beta)^2 \;.
\end{equation}
This expression will be used later on in this paper to compare this theoretical model with data from our experiment.\\

\section{Experiment}

\begin{figure}
\begin{center}
\includegraphics[scale=0.7]{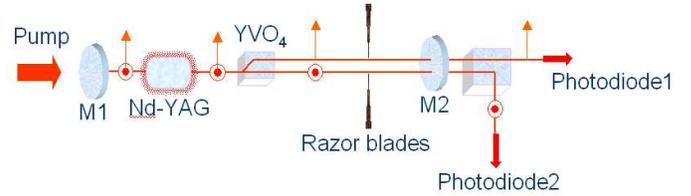}
\end{center}
\caption{Sketch of our experimental laser cavity setup. Each razor blade can be translatory moved perpendicularly to the
laser axis, in order to create additional losses to the corresponding laser mode. M1 and M2 are the cavity
mirrors.}\label{figure 0}
\end{figure}

We now turn to the description of our experimental setup, sketched in figure~\ref{figure 0}. We use a 18-cm-long linear
laser cavity with a 2.5-cm-long Nd-YAG crystal as a gain medium. The cavity also contains a 10-mm-long uniaxial
birefringent crystal (YVO$_4$) cut at 45~deg of its optical axis, in order to spatially separate the two orthogonal
modes of the cavity. We have checked that the two perpendicular ordinary and extraordinary polarizations correspond to
the two spatially separated modes inside the cavity and that no significant cross-coupling inducing forked eigenstate
operation \cite{fork} was induced inside the Nd-YAG crystal (which imposes a specific crystal orientation in the
$\langle 100 \rangle$ case, see further). We have also checked, using a Fabry-Perot analyzer, that each one of the two
perpendicular laser modes was longitudinally single-mode. This is probably due to the fact that the YVO$_4$ crystal acts
as an etalon and also due to the near-threshold operation of the laser.\\

The razor blades, placed in the vicinity of the two separated beams, are intended for creating additional losses to the
corresponding laser modes (i.e. changing $\gamma_1$ and $\gamma_2$). We limit the losses introduced by the razor blade
to about $1\%$, allowing us to neglect the beam truncation and to keep a good overlap of the beams in the active medium.
The intensities of the two orthogonal modes are then monitored on two photodiodes after being separated by a polarizing
beam splitter. Typically, the position of one razor blade is changed periodically in time (using a piezoelectric
transducer), and the relative dependence of $I_1$ and $I_2$ is monitored on an oscilloscope. We successively introduce
razor blade losses to the $1$ and $2$ modes, thus obtaining experimental values for $(\Delta I_2/\Delta
\gamma_1)/(\Delta I_1/\Delta \gamma_1)$ and $(\Delta I_1/\Delta \gamma_2)/(\Delta I_2/\Delta \gamma_2)$. The
coupling constant is eventually deduced from expression (\ref{def}). \\

It is worth noticing that this result is independent, at least in the framework of our theoretical model, of most laser
parameters, in particular saturation intensities $I_s^1$ and $I_s^2$, pumping rate and calibration of photodiodes. The
dominant error source in this measurement is intensity self-modulation, which introduces an uncertainty in the slope
measurement. Another error source is residual pumping anisotropy. In order to assess the contribution of the latter
effect, we have checked experimentally that pumping light had no preferential polarization axis with an accuracy better
than a few percents, resulting in a relative error of a few percents on the measurement of $C$. The overall measurement
error is estimated to $\delta C / C \simeq 10\%$.

\subsection{Case of the $\langle 111 \rangle$ Nd-YAG}

\begin{figure}
\begin{center}
\includegraphics[scale=0.52]{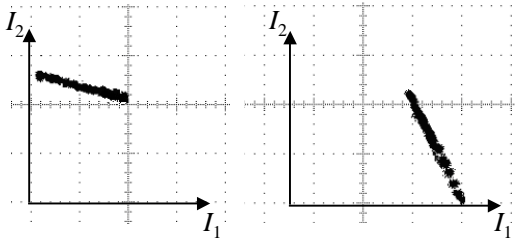}
\end{center}
\caption{Experimental curves showing $I_2$ versus $I_1$ in the $\langle 111 \rangle$ case when the loss rates $\gamma_1$
(left curve) and $\gamma_2$ (right curve) are changed. Slopes measurements provide the following values~: $(\Delta
I_2/\Delta \gamma_1)/(\Delta I_1/\Delta \gamma_1) = -0.3 \pm 0.03$ (left curve) and $(\Delta I_1/\Delta
\gamma_2)/(\Delta I_2/\Delta \gamma_2) = -0.5 \pm 0.05$, leading to $C = 0.15 \pm 0.015$.}\label{figure 1}
\end{figure}

In a first experiment, we use a typical $\langle 111 \rangle$ Nd-YAG crystal as a gain medium, with arbitrary
orientation. The raw output light from the fibred laser diode is focused on the crystal for optical pumping. Following
the method described above for the measurement of the coupling constant, we have obtained the data reported on
figure~\ref{figure 1}. This leads to the following value~: $C \simeq 0.15 \pm 0.015$, in very good agreement with the
previously-published value $C \simeq 0.16 \pm 0.03$ \cite{apl_brunel}. This agreement is probably owing to the fact that
the coupling constant is relatively independent of most laser parameters, as pointed out previously, provided the two
cavity modes are properly separated by the YVO$_4$ crystal and spatial hole burning can be neglected (which can be shown
to be the case here and in the work of reference \cite{apl_brunel}). Taking into account this new experimental value
(with smaller error bars than in \cite{apl_brunel}), the following finer estimate of the ellipticity $\beta$ can be
deduced from equation (\ref{Cwithbeta})~: $\beta \simeq 2.2 \%$ ($\pm 0.2\% $).

\subsection{Case of the $\langle 100 \rangle$ crystal}

\begin{figure}
\begin{center}
\includegraphics[scale=0.6]{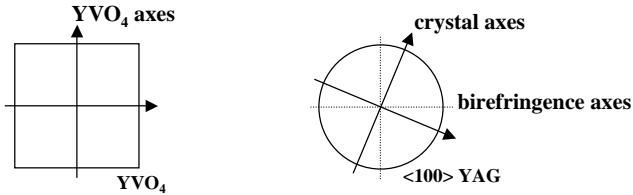}
\end{center}
\caption{Relative orientation, around the longitudinal cavity axis, of the YVO$_4$ crystal (left) and the $\langle 100
\rangle$ Nd-YAG crystal (right). As can be seen on this sketch, the YAG crystal has been oriented in order to align its
residual birefringence axes with the YVO$_4$ crystal axes, in order to ensure that the cavity eigenmodes are linearly
polarized along the YVO$_4$ axes (instead of being forked modes).}\label{figure axes}
\end{figure}

\begin{figure}
\begin{center}
\includegraphics[scale=0.52]{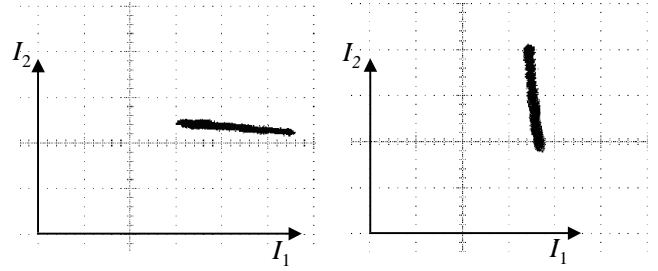}
\end{center}
\caption{Experimental curves showing $I_2$ versus $I_1$ in the $\langle 100 \rangle$ case (angle between laser and
crystal axes~: 10$\pm$1~deg)when the loss rates $\gamma_1$ (left curve) and $\gamma_2$ (right curve) are changed. Slopes
measurements provide the following values~: $(\Delta I_2/\Delta \gamma_1)/(\Delta I_1/\Delta \gamma_1) = -0.11 \pm 0.01$
(left curve) and $(\Delta I_1/\Delta \gamma_2)/(\Delta I_2/\Delta \gamma_2) = -0.1 \pm 0.01$, leading to $C = 0.011 \pm
0.001$.}\label{figure 2}
\end{figure}

In a second experiment, the $\langle 111 \rangle$ gain medium is replaced by a $\langle 100 \rangle$ Nd-YAG crystal,
manufactured by the German company FEE GmbH. The crystallographic axes are known precisely by X-ray analysis. A small
birefringence is observed for this crystal, probably due to mechanical stress from the mount. In order to avoid the
appearance of forked modes in the laser cavity \cite{fork}, it is necessary to properly align the birefringence axis of
the YAG crystal with the ordinary and extraordinary polarization axis of the YVO$_4$ crystal. As illustrated on figure
\ref{figure axes}, such a configuration is experimentally obtained, on our setup, with an angle $\alpha=10\pm 1$~deg
between the crystal axes and the YVO$_4$ axes. With such an alignment, the cavity eigenmodes are linearly polarized and
coincide with the ordinary and extraordinary polarization axis of the YVO$_4$ crystal, making the experiment suitable
for comparison with our theoretical model. It is then possible to measure the coupling constant between both orthogonal
modes following the previously-described method, as reported on figure~\ref{figure 2}. The result is $C \simeq 0.011 \pm
0.001$, to be compared with the theoretical value predicted by our model (equation (\ref{expdeCfinal}) with
$\alpha=10\pm 1$~deg and $\beta \simeq 2.2 \%\pm 0.2\%$), namely $C \simeq 0.011\pm 0.001$. This remarkable agreement is
an experimental evidence for the simple theoretical description proposed in this paper.\\

\section{Conclusion}

To summarize, we have proposed a simple yet accurate theoretical model for describing the orientation of transition
dipoles in a Nd-YAG crystal, by making the following hypotheses~: first, dipoles are slightly elliptic; second, their
main directions are collinear with the crystallographic axes. This model has been tested by making experimental
measurements of the coupling constant between two orthogonal modes of a Nd-YAG laser for two different crystal cuts
($\langle 111 \rangle$ and $\langle 100 \rangle$). A remarkable quantitative agreement between theory and experiment has
been observed. In particular, this study is an experimental evidence for the fact that transition dipoles are indeed
aligned with crystallographic axes of the Nd-YAG crystal, with an accuracy better than 1~deg.\\

The significant reduction of coupling between orthogonally polarized modes in a laser cavity using a $\langle 100
\rangle$-cut gain medium, as demonstrated in this paper, could be used to significantly increase the stability of
bi-frequency lasers, with possible applications in the field of lidar or multioscillator ring laser gyroscopes. Our work
predicts that the more favorable situation for such applications will occur when the crystallographic axes are aligned
with the laser cavity axes, with a minimum achievable coupling constant as small as $4 \beta^2 \simeq 2.5\,10^{-3}$.
Furthermore, the simple and original protocol proposed in this paper could be applied to other kinds of solid-state gain
media or saturable absorbers, in order to probe the orientation of their active dipoles.\\

\begin{acknowledgments}
The authors are happy to thank Philippe Goldner from Chimie ParisTech and Daniel Rytz from FEE GmbH for helpful
discussions.
\end{acknowledgments}

\section{Appendix: Calculation of the coupling constant for an arbitrary crystal orientation in the $\langle 111 \rangle$
case without dipole ellipticity}

In this appendix, we will show that in the case of a $\langle 111 \rangle$ crystal with linear dipoles oriented along
the crystallographic axes, the coupling constant is independent of the orientation between the crystal axes and the
directions of the laser modes. To this end, we consider linear dipoles oriented along the crystallographic axes,
namely~:
\begin{equation} \label{dipol2}
\mathbf{u}_1= \left( \begin{tabular}{c} 1 \\ 0 \\ 0 \end{tabular} \right) \quad\!\! \textrm{,} \quad \mathbf{u}_2=
\left(
\begin{tabular}{c} 0 \\ 1 \\ 0 \end{tabular} \right) \quad\! \textrm{and} \quad\! \mathbf{u}_3= \left(
\begin{tabular}{c} 0 \\ 0 \\ 1 \end{tabular} \right) \;.
\end{equation}
Any pair of orthogonal directions for the laser modes in the transverse plane can be obtained by rotating (by the
appropriate angle $\alpha$) the initial base defined by (\ref{basis}), which reads~:
\begin{equation} \label{modes2} \left\{
\begin{split} & \mathbf{x}_1=\frac{\cos \alpha}{\sqrt{2}} \left( \begin{tabular}{c} 1 \\ -1 \\ 0 \end{tabular} \right) +
\frac{\sin \alpha}{\sqrt{6}} \left( \begin{tabular}{c} 1 \\ 1 \\ -2 \end{tabular} \right) \;, \\
& \mathbf{x}_2=\frac{-\sin \alpha}{\sqrt{2}} \left( \begin{tabular}{c} 1 \\ -1 \\ 0 \end{tabular} \right) + \frac{\cos
\alpha}{\sqrt{6}} \left( \begin{tabular}{c} 1 \\ 1 \\ -2 \end{tabular} \right) \;.
\end{split} \right.
\end{equation}
A straightforward although tedious calculation leads, from equations (\ref{dipol2}) and (\ref{modes2}), to the following
expression~:
\begin{equation} \label{A1}
\sum_{i=1}^3 \cos^2 \left(\widehat{\mathbf{x}_1,\mathbf{u}_i} \right)\cos^2 \left(\widehat{\mathbf{x}_2,\mathbf{u}_i}
\right) = \frac{1}{6} \;.
\end{equation}
Similarly, we obtain~:
\begin{equation} \label{A2}
\sum_{i=1}^3 \cos^4 \left(\widehat{\mathbf{x}_1,\mathbf{u}_i} \right) = \frac{1}{2} \;\;\, \textrm{and} \;\;
\sum_{i=1}^3 \cos^4 \left(\widehat{\mathbf{x}_2,\mathbf{u}_i} \right) = \frac{1}{2} \,.
\end{equation}
It is a remarkable fact that expressions (\ref{A1}) and (\ref{A2}) are independent of $\alpha$. Using expression
(\ref{C}) for the coupling constant eventually leads to $C=1/9$, independently of the orientation between the $\langle
111 \rangle$ crystal and the laser modes.

\end{document}